\documentclass[prb,twocolumn,amsmath,amssymb,showpacs]{revtex4}
\usepackage{amsmath}
\usepackage{pstricks}
\usepackage{xspace}
\usepackage{comment}
\usepackage{graphicx}
\usepackage[english]{babel}
\usepackage{bm}
\usepackage{ulem}
\usepackage{color}%
\usepackage{amsfonts}%
\usepackage{amssymb}
\begin{document}

\title{Strongly anisotropic roughness in surfaces driven by an oblique particle flux}
\author{B. Schmittmann}
\affiliation{Center for Stochastic Processes in Science and Engineering, \\
Department of Physics, Virginia Tech, Blacksburg, VA 24061-0435, USA}
\email{schmittm@vt.edu}
\author{Gunnar Pruessner}
\affiliation{Department of Physics (CMTH), Imperial College, London SW7 2BW, UK}
\author{Hans-Karl Janssen}
\affiliation{Institut f\"{u}r Theoretische Physik III, Heinrich-Heine-Universit\"{a}t,
40225 D\"{u}sseldorf, Germany}
\date{January 16, 2006}

\begin{abstract}
Using field theoretic renormalization, an MBE-type growth process with an
obliquely incident influx of atoms is examined. The projection of the beam on
the substrate plane selects a ``parallel'' direction, with rotational
invariance restricted to the transverse directions. Depending on the behavior
of an \textit{effective} anisotropic surface tension, a line of second order
transitions is identified, as well as a line of potentially first order
transitions, joined by a multicritical point. Near the second order
transitions and the multicritical point, the surface roughness is strongly
anisotropic. \textit{Four} different roughness exponents are introduced 
and computed, describing the surface in different directions, 
in real or momentum space. The results presented challenge an 
earlier study of the multicritical point.

\end{abstract}
\maketitle

\section{Introduction}

The fabrication of numerous nanoscale heterostructures requires the controlled
deposition of material onto a substrate. A variety of deposition processes are
used, depending on desired surface structure and device performance. Molecular
beam epitaxy (MBE), involving directed beams of incident atoms, is
particularly suitable if lower growth temperatures and precise in situ control
and characterization are desired \cite{Joyce:1985}. It is an important goal of
both theoretical and experimental investigations to gain an understanding of
the resulting surface morphology, in terms of its spatial and dynamic
height-height correlations, or more specifically, its roughness.

Beyond the obvious implications for nanoscale devices, surface growth problems
also constitute an important class of generic nonequilibrium phenomena
\cite{Krug:1997}. Particles are deposited on the surface and may diffuse
around on it. If deposition occurs from a vapor, desorption or bulk defect
formation tend to be important processes; in contrast, both mechanisms can
often be neglected in MBE (see for example \cite{Lagally:1990}). After an
initial transient, a steady state is established which is characterized by
time-independent macroscopic properties, provided a suitable reference frame
is chosen. Generically, detailed balance is broken by the incident particle
flux \cite{SiegertPlischke:1992}, so that this steady state cannot be
described by a Boltzmann distribution; instead, its statistical properties
have to be determined directly from its dynamical evolution. If one is
primarily interested in universal, large scale, long time characteristics, the
dynamical evolution can often be cast as a Langevin equation which can be
analyzed using techniques from renormalized field theory.

Here, we extend a model \cite{MarsiliETAL:1996} due to Marsili \textit{et al.}
to describe MBE-type or ballistic deposition processes with oblique particle
incidence. Focusing on large scale properties such as surface roughness, we
exploit a coarse-grained (continuum) approach. Adopting an idealized
description \cite{WolfVillain:1990,Villain:1991,LaiDasSarma:1991,Janssen:1996}%
, particle desorption and bulk defect formation will be neglected so that all
(deterministic) surface relaxation processes are mass-conserving, i.e., can be
written as the gradients of equilibrium and non-equilibrium currents. Shot
noise in the deposition process requires the addition of a stochastic term to
the growth equation. Since the particle beam selects a preferred
(``parallel'') direction in the substrate plane, the resulting Langevin
equation is necessarily anisotropic. The interplay of interatomic interactions
and kinetic effects, such as Schwoebel barriers, generates an anisotropic
effective surface tension which can become very small or even vanish. Due to
the anisotropy, this leads to four different regimes with potentially
scale-invariant behaviors. We analyze these four regimes, identify the
scale-invariant ones, and compute the associated anisotropic roughness 
exponents.

Models with oblique particle incidence have been investigated previously.
Focusing on vapor-deposited thin films, Meakin and Krug
\cite{Meakin:1988,MeakinKrug:1990,MeakinKrug:1992} considered the ballistic
deposition of particles under near-grazing incidence. Under these conditions,
columnar patterns form which shield parts of the growing surface from incoming
particles. The large scale properties of these structures can be characterized
in terms of anisotropic scaling exponents, differentiating parallel and
transverse directions \cite{KrugMeakin:1989,MeakinKrug:1992}.

Following Marsili et al.\cite{MarsiliETAL:1996}, our model differs from 
Krug and Meakin's approach in two
important respects. First, surface overhangs and shadowing effects are
neglected so that our results are restricted to near-normal
incidence. Second, our model is designed for ``ideal MBE''-type growth so that
it falls outside the Kardar-Parisi-Zhang universality class
\cite{KardarParisiZhang:1986} for mass non-conserving growth. However, 
extending the work of Marsili \textit {et al} \cite{MarsiliETAL:1996}, 
we include a possibly anisotropic
\textit{effective} surface tension and investigate its effects systematically
(see further comments in the following section). Due to the anisotropy, this
contribution actually consists of two terms, one controlling relaxation
parallel and the other transverse to the beam direction, with coupling
constants $\tau_{\Vert}$ and $\tau_{\bot}$, respectively. Depending on which
of these two couplings, $\tau_{\Vert}$ or $\tau_{\bot}$, vanishes first,
ripple-like surface structures are expected, aligned transverse to the soft
direction. The
roughness properties near these two instabilities, characterized respectively
by $\tau_{\bot}=0$ while $\tau_{\Vert}>0$, and $\tau_{\bot}>0$ while
$\tau_{\Vert}=0$, are discussed in this paper for the first time.

The original theory of Marsili \textit {et al}\cite{MarsiliETAL:1996} 
is recovered only if both
couplings, $\tau_{\Vert}$ \textit{and} $\tau_{\bot}$, vanish simultaneously.
Since the latter requires the careful tuning of \textit{two} parameters, it is
much less likely to be experimentally relevant than either of the other two
instabilities in which only a \textit{single} parameter must be adjusted.
Referring all technical details to a separate publication \cite{SPJ-Helsinki},
we will point out very briefly that the fixed point and the roughness
exponents reported in Marsili \textit {et al}\cite{MarsiliETAL:1996} must 
be significantly revised.

An important aspect discussed in this article is how to translate surface data
into roughness exponents, for inherently anisotropic surface models such as
the one analyzed here. If we generalize the standard definitions familiar from
isotropic problems, we arrive at \textit{four different} roughness exponents.
\cite{SchmittmannZia:1995} Two of
these characterize real-space scans along and transverse to the beam
direction, and the remaining two are needed to describe scattering (i.e.
momentum space) data with parallel or transverse momentum transfer. All four
are related by simple scaling laws, but possess distinct numerical values.
When analyzing experimental data, it is therefore essential to be aware of
these subtleties.

The paper is organized as follows. We first present the underlying Langevin
equation for a single-valued height field and briefly review the physical
origin of its constituents. We then present a careful definition of the
roughness exponents, based on height-height correlation functions and
their structure
factors. Turning to the renormalization group (RG) analysis, we first discuss
a simple scaling symmetry of our model which allows us to identify a set of
effective coupling constants. The invariance of our model with respect to
\textit{tilts} of the surface is much more powerful. Since this symmetry is
continuous, it gives rise to a Ward identity which relates different vertex
functions. This simplifies the renormalization procedure considerably. We then
present our main results for the scaling properties of correlation and
response functions for the four different cases: (o) both $\tau_{\Vert}$ and
$\tau_{\bot}$ are positive; (i) $\tau_{\Vert}$ remains positive while
$\tau_{\bot}$ vanishes; (ii) $\tau_{\Vert}$ vanishes while $\tau_{\bot}$
remains positive; and finally, (iii) both $\tau_{\Vert}$ and $\tau_{\bot}$
vanish simultaneously. Roughness and dynamic critical exponents are derived.
We conclude with a short summary and a discussion of the experimental evidence.

\section{The model}

\begin{figure}[ptb]
\scalebox{0.75}{\includegraphics{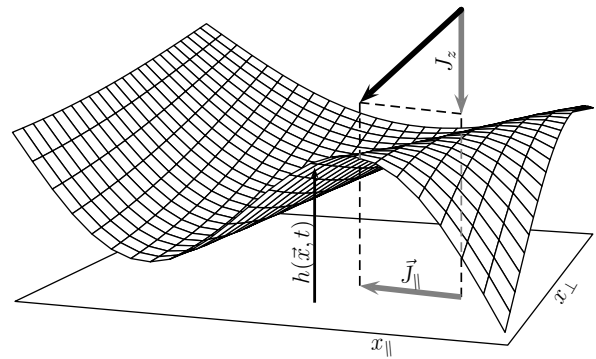}}
\caption{ Cartoon of the arrangement of surface and incoming flux. The field
$h(\mathbf{x},t)$ is the height of the interface over the substrate. Its
coordinate system can always be chosen such that $x_{\parallel}$ is parallel
to the projection of the flux on the substrate. The remaining component of the
latter is the (negative) flux $J_{z}$. \label{fig:fig_growth}}%
\end{figure}

We focus on long time, large distance phenomena of the growing
surface. Under suitable conditions \cite{MarsiliETAL:1996}, surface overhangs
and shadowing effects may be neglected, so that the surface can be described
by a single-valued height field, $h(\mathbf{r},t)$, where $\mathbf{r}$ denotes
a $d$-dimensional vector in a reference (substrate) plane, the $z$-axis is
normal to that plane, and $t$ denotes time, see Fig.~\ref{fig:fig_growth}. The
no-overhang assumption can be justified \textit{a posteriori} if the
calculated interface roughness exponent are found to be less than unity. The
time evolution of the interface is described by a Langevin equation of the
form
\begin{equation}
\partial_{t}h=G[h]+\eta\label{generic_Langevin}%
\end{equation}
Here, $\eta$ denotes the effects of shot noise, and $G[h]$ models the
deterministic part of the surface evolution, assumed to be mass-conserving in
a suitable coordinate system so that $G[h]$ can be written as a divergence,
$G[h]=\mathbf{\nabla\cdot F}[h]$. One contribution to $G[h]$ is due to the
incident flux; the other contribution arises from surface diffusion. All of
these contributions can be derived using the principle of re-parametrization
invariance \cite{MarsiliETAL:1996b} and are discussed in the following.

\begin{figure}[ptb]
\scalebox{0.75}{\includegraphics{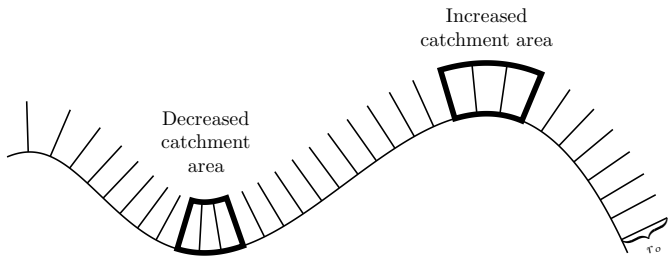}}
\caption{ Schematic representation of the growth inhibition and amplification
by either finite adatom radii ($r_o$) or steering. In both cases, the
flux contributing to growth at some point on the surface is to be measured a
distance $r_o$ normal to the surface. The left box shows the reduced effective
surface in valleys, the right one the corresponding effect 
on peaks\cite{MazorETAL:1988,Krug:2005}. 
\label{fig:surface_shift} }
\end{figure}

As shown in Fig.~\ref{fig:fig_growth}, the incident particle current has a
normal component $J_{z}$ and a component parallel to the substrate plane,
$\mathbf{J}_{\parallel}$. In the following, the co-ordinate system is rotated
such that one of the axes, labelled $x_{\Vert}$, is aligned with
$\mathbf{J}_{\parallel}$. The particles themselves are of finite size with
radius $r_{o}$, which implies that the flux responsible for growth at some
point on the surface is to be measured at a distance $r$ normal to the
surface, see Fig.~\ref{fig:surface_shift}. This effect has been discussed in
detail in the literature, see
\cite{LeamyGilmerDirks:1980,MazorETAL:1988,MarsiliETAL:1996b}. Neglecting
higher order terms in the spirit of a gradient expansion, to leading order,
this effect gives rise to a deterministic term of the form
\begin{equation}
G_{\mathrm{drive}}[h]=-J_{z}+\mathbf{J}_{\parallel}\cdot\mathbf{\nabla}%
h+r_{o}J_{z}\nabla^{2}h-r_{o}\left(  \mathbf{J}_{\parallel}\cdot
\mathbf{\nabla}h\right) \nabla^{2}h \label{music}
\end{equation}
Most remarkably, so called ``steering'' leads to the same terms in leading
order \cite{Krug:2005}. Steering implies that deposited atoms are 
deflected towards the
surface normal as soon as they reach a certain distance above it, due to an
attractive force exerted by the particles in the deposit. 
Returning to Eq.~(\ref{music}), we note that the first two terms in 
$G_{\mathrm{drive}}[h]$ can be removed by a Galilei 
transformation $h(\mathbf{r},t)\rightarrow
h(\mathbf{r+J}_{\parallel}t,t)-J_{z}t$. From now on, we always work in
this co-moving frame.

In addition to being driven by the incident flux, Eq.~(\ref{music}), the 
surface also relaxes via diffusion \textit{along} the surface, leading
to a quartic term of the form \cite{Mullins:1963}
\begin{equation}
G_{\text{\textrm{relaxation}}}\ =-\mu\nabla^{4}h \label{diff}%
\end{equation}
In the following, terms of the form $\tau\nabla^{2}h$ play a particularly
important role, since they determine which of the various critical regimes can
be accessed. Eq.~(\ref{music}) already contains such a term, induced by 
$J_{z}$, but other contributions of this type are possible, 
e.g., a negative term from a step edge
(Schwoebel) barrier \cite{SchwoebelShipsey:1966,Schwoebel:1969,Villain:1991}
or a positive one due to a surface tension. Moreover, even if such a term
were initially absent, it would actually be 
generated under renormalization group transformations and is therefore
intrinsically present. In contrast to Marsili et al.\cite{MarsiliETAL:1996}, 
we include it from the very beginning.

It is essential to note that the nonlinear term in Eq.~(\ref{music}),
\[
r_{o}\left(  \mathbf{J}_{\parallel}\cdot\mathbf{\nabla}h\right)  \,\nabla
^{2}h\equiv\lambda\left(  \partial_{\parallel}h\right)  \,\nabla^{2}h
\]
introduces an anisotropy into the system which breaks the full rotational
symmetry within the $d$-dimensional space of the substrate. As a consequence,
there is no reason to expect isotropic coupling constants (such as $\mu$ or
$\tau$) for the linear contributions. Instead, any coarse-graining of the
microscopic (atomic level) theory is expected to give rise to different
couplings (such as $\tau_{\parallel}$ and $\tau_{\perp}$)\ in the parallel and
perpendicular subspaces, and this is indeed confirmed by the renormalization
group. If these anisotropies are incorporated into the theory, 
preserving only rotational
invariance in the $(d-1)$-dimensional transverse subspace, the Langevin
equation (\ref{generic_Langevin}) takes the form

\begin{align}
\gamma^{-1}\partial_{t}h &  = \tau_{\parallel}\partial_{\parallel}^{2}%
h+\tau_{\perp}\nabla_{\perp}^{2}h-\mu_{\parallel}\partial_{\parallel}%
^{4}h-2\mu_{\times}\nabla_{\perp}^{2}\partial_{\parallel}^{2}h \nonumber \\
& -\mu_{\perp}(\nabla_{\perp}^{2})^{2}h+\left(  \partial_{\parallel}h\right)%
  \,\bigl(
\lambda_{\parallel}\partial_{\parallel}^{2}h+\lambda_{\perp}\nabla_{\perp}%
^{2}h\bigr)  +\eta\nonumber\\
&  =-\partial_{\parallel}j_{\parallel}-\mathbf{\nabla}_{\perp}\mathbf{j}%
_{\perp}+\eta\label{Langevin}%
\end{align}
where we have introduced an explicit time scale for convenience. The surface
currents are given by
\begin{align}
j_{\parallel}  = & -\partial_{\parallel}\bigl(  \tau_{\parallel}%
h-\mu_{\parallel}\partial_{\parallel}^{2}h-\mu_{\times}\nabla_{\perp}%
^{2}h\bigr) \nonumber \\  
& -\frac{\lambda_{\parallel}}{2}\bigl(  \partial_{\parallel
}h\bigr)  ^{2}+\frac{\lambda_{\perp}}{2}\bigl(  \nabla_{\perp}h\bigr)
^{2}\,,\nonumber\\
\mathbf{j}_{\perp} = & -\nabla_{\perp}\bigl(  \tau_{\perp}h-\mu_{\perp}%
\nabla_{\perp}^{2}h-\mu_{\times}\partial_{\parallel}^{2}h\bigr) \nonumber \\
& -\lambda_{\perp}\bigl(  \nabla_{\perp}h\bigr)  \,\bigl(  \partial_{\parallel
}h\bigr)  \,. \label{curr}%
\end{align}
The scalar differential operator $\partial_{\parallel}$ operates only along
$x_{\parallel}$ (see Fig.~\ref{fig:fig_growth}), while the vector
$\mathbf{\nabla}_{\bot}$ operates in the $(d-1)$ dimensional subspace
perpendicular to $x_{\parallel}$. We also note that the nonlinearity now
splits into two distinct terms, with couplings $\lambda_{\parallel}$ and
$\lambda_{\perp}$, respectively. Finally, the conserved nature of the
deterministic surface evolution is displayed explicitly here.

The randomness of particle aggregation on the surface is captured by the
non-conserved white noise $\eta(\mathbf{r},t)$ with zero average and second
moment
\begin{equation}
\left\langle \eta(\mathbf{r},t)\eta(\mathbf{r}^{\prime},t^{\prime
})\right\rangle =2\gamma^{-1}\delta(\mathbf{r}-\mathbf{r}^{\prime}%
)\ \delta(t-t^{\prime})\ . \label{noise_correl}%
\end{equation}

Eq.~(\ref{Langevin}) forms the basis for the following analysis. Its
properties are controlled by the dominant terms in the gradient expansion. To
ensure the stability of the linear theory, we require 
$\mu_{\parallel},\mu_{\perp} > 0$, and 
$\mu_{\times}\geq-(\mu_{\parallel}\mu_{\perp})^{1/2}$. The two
couplings $\tau_{\parallel}$ and $\tau_{\perp}$ play the role of critical
control parameters. If both are positive, the nonlinearities become
irrelevant, and the problem reduces to the well-known Edwards-Wilkinson
equation \cite{EdwardsWilkinson:1982}. In contrast, if one or both of them
vanish, the long time, long distance properties of the theory change
dramatically: The surface undergoes an instability and forms characteristic
spatial patterns. If just one of the couplings goes soft,
these patterns take the form of ripples (similar to corrugated roofing) 
transverse to the soft direction. If both couplings become
negative, the surface develops mounds or ``wedding cakes'' \cite{Krug:1997}.
Focusing only on the onset of these instabilities, four different cases emerge
whose properties are discussed in the following: (o) the ``disordered'' 
phase, corresponding to the linear theory with $\tau_{\parallel}>0$, 
$\tau_{\perp}>0$; (i) a line of continuous
transitions $\tau_{\parallel}>0$, $\tau_{\perp}\rightarrow0$; (ii) a line of
\textit{possibly first} order transitions $\tau_{\parallel}\rightarrow0$,
$\tau_{\perp}>0$; and (iii) the multicritical (critical end-) point
$\tau_{\parallel}\rightarrow0$, $\tau_{\perp}\rightarrow0$.
Before we turn to any
technicalities, however, we first discuss an important physical issue, namely,
the definition of appropriate roughness exponents.

\section{Anisotropic roughness exponents}

The roughness of the surface, and the associated roughness exponents, if they
exist, are easily measured experimentally. They can be determined from
real-space images of the surface or from scattering data in momentum space.
Within our theoretical framework, roughness exponents can be extracted from
the height-height correlation function,
\begin{equation}
C(\mathbf{r}-\mathbf{r}^{\prime},t-t^{\prime})\equiv\left\langle
h(\mathbf{r},t)h(\mathbf{r}^{\prime},t^{\prime})\right\rangle
\label{def_corrfct}%
\end{equation}
Since we focus on the steady state in the absence of spatial boundaries, we
assume translational invariance in space and time. The spatial Fourier transform 
of $C$ is the dynamic structure factor,
\begin{equation}
C(\mathbf{q},t)=\int d^{d}r\,C(\mathbf{r},t)\,e^{i \mathbf{q}%
\cdot\mathbf{r}}\,. \nonumber 
\end{equation}
In the absence of anisotropies, the asymptotic scaling behavior of
Eq.~(\ref{def_corrfct}) can be written in the form
\begin{equation}
C(\mathbf{r},t)=\left|  \mathbf{r}\right|  ^{2\chi}c(t/\left|  \mathbf{r}%
\right|  ^{z}) \label{isotropic_realspace_roughness}%
\end{equation}
where $\chi$ denotes the roughness exponent and $z$ the dynamic exponent of
the surface while $c$ is a universal scaling function. In Fourier space, the
behavior of $C(\mathbf{r},t)$ translates into
\begin{equation}
\widetilde{C}(\mathbf{q},t)=\left|  \mathbf{q}\right|  ^{-(d+2\chi
)}\widetilde{c}\left(  t\left|  \mathbf{q}\right|  ^{z}\right)
\label{isotropic_kspace_roughness}%
\end{equation}

In the presence of strong anisotropy, where the \textit{scaling} of the
correlation function depends on the direction, the situation becomes more
complex. Anticipating some of the following results, a key finding of the
present article is the existence of four \textit{different} roughness
exponents, characterizing the surface along the parallel or transverse
directions, in real or in momentum space. While they are directly related by
scaling laws, it is essential to realize that they take different 
\textit {numerical values}. 
In the interpretation of actual experimental data, it is therefore
important to identify the appropriate member of this set of four exponents in
order to compare with theoretical predictions.

In our scaling analysis below, we adopt the conventional exponent definitions
from critical dynamics. The exponent $\nu$ controls the divergence of the
correlation length, while $\eta$ denotes the anomalous dimension of the height
field which appears in all correlation functions. Due to the presence of
anisotropy, an additional exponent, i.e., the strong anisotropy exponent
$\Delta$, is required to reflect the different scaling of distances or wave
vectors in different directions \cite{SchmittmannZia:1995}. If $l$ denotes an
arbitrary transverse momentum scale, so that $\left|  \mathbf{q}_{\bot
}\right|  \propto l$, we introduce $\Delta$ via $q_{\Vert}\propto l^{1+\Delta
}$. With this definition, one finds that, asymptotically, the structure factor
is a generalized homogeneous function of its variables
\begin{equation}
\widetilde{C}(q_{\Vert},\mathbf{q}_{\bot};t)=l^{-4+\eta}\widetilde
{C}(q_{\Vert}/l^{1+\Delta},\mathbf{q}_{\bot}/l,l^{z}t) \label{SF_MS}%
\end{equation}
and in real space, the two-point function takes the form
\begin{equation}
C(x_{\Vert},\mathbf{r}_{\bot};t)=l^{d+\Delta-4+\eta}C(l^{1+\Delta}x_{\Vert
},l\mathbf{r}_{\bot},l^{z}t) \label{G_RS}%
\end{equation}
In analogy to Eq.~(\ref{isotropic_realspace_roughness}) two roughness
exponents are defined in real space, $\chi_{\bot}$ and $\chi_{\Vert}$, via
\begin{align}
C(0,\mathbf{r}_{\bot};t)  &  \equiv\left|  \mathbf{r}_{\bot}\right|
^{2\chi_{\bot}}c_{\bot}(t/\left|  \mathbf{r}_{\perp}\right|  ^{z}) \nonumber \\
C(x_{\Vert},\mathbf{0};t)  &  \equiv\left|  x_{\Vert}\right|  ^{2\chi_{\Vert}%
}c_{_{\Vert}}\bigl(  t/x_{\Vert}^{z/(1+\Delta)}\bigr) \label{aniso_RS}
\end{align}
Of course, this is only meaningful if the two scaling functions $c_{\bot}$ and
$c_{_{\Vert}}$ approach finite and non-zero constants when their arguments
vanish. Under this assumption, the two exponents
\begin{eqnarray}
\chi_{\bot} &=&\frac{1}{2}\bigl[  4-\left(  d+\Delta\right)  -\eta\bigr] 
\nonumber \\
\chi_{\Vert} &=&\frac{1}{2}(1+\Delta)^{-1}%
\bigl[ 4-\left(d+\Delta\right) -\eta\bigr] \label{Chi_RS}%
\end{eqnarray}
are read off immediately. In order to define the corresponding exponents in
momentum space, $\tilde{\chi}_{\bot}$ and $\tilde{\chi}_{\Vert}$, we focus on
two structure factors which are easily accessible experimentally,
especially if we set $t=0$:
\begin{eqnarray}
\widetilde{C}(0,\mathbf{q}_{\bot},t)  &  \equiv\left|  \mathbf{q}_{\bot
}\right|^{-(d+2\tilde{\chi}_{\bot})}\widetilde{c}_{\bot}\left(
t\left|  \mathbf{q}_{\bot}\right|  ^{z}\right) \nonumber \\
\widetilde{C}(q_{\Vert},\mathbf{0},t)  &  \equiv q_{\Vert}^{-(d+2\tilde
{\chi}_{\Vert})}\widetilde{c}_{_{\Vert}}\bigl( t\,q_{\Vert}%
^{z/(1+\Delta)}\bigr) \label{aniso_MS}
\end{eqnarray}
Again, provided the scaling functions $\widetilde{c}_{\bot}$ and
$\widetilde{c}_{\Vert}$ are nonsingular and non-zero in the limit of vanishing
argument, one reads off
\begin{eqnarray}
\tilde{\chi}_{\bot} &=&\frac{1}{2}\bigl[  4-d-\eta\bigr] \nonumber \\ 
\tilde{\chi}_{\Vert} &=&\frac{1}{2}\bigl[\frac{4-\eta}{1+\Delta}-d \bigr]
\label{Chi_MS}%
\end{eqnarray}
The key observation is that $\tilde{\chi}_{\bot}=\chi_{\bot}$ and $\tilde
{\chi}_{\Vert}=\chi_{\Vert}$ only if the anisotropy exponent $\Delta$
vanishes. Thus, contrary to Eqs.~(\ref{isotropic_realspace_roughness}) and
(\ref{isotropic_kspace_roughness}) which are equivalent definitions of the
roughness exponent in isotropic systems, the corresponding definitions for
anisotropic systems will typically give rise to different exponents.

In the following, we explicitly compute the scaling exponents in the previous
expressions, and also confirm the underlying scaling form. To unify the
discussion, we first recast the Langevin equation as a dynamic field theory,
collect the elements of perturbation theory, and then identify the upper
critical dimensions and marginal nonlinearities for the four cases. Our final
goal is a systematic derivation of the scaling properties of correlation and
response functions.

\bigskip

\section{Renormalization group analysis}

\subsection{Power counting and mean-field exponents.}

In this section, we assemble the basic components of the field-theoretic
analysis, leaving technical details to \cite{SPJ-Helsinki}. The formalism
becomes most elegant if we introduce a response field $\tilde{h}%
(\mathbf{r},t)$ and recast the Langevin equation (\ref{Langevin})\ as a
dynamic functional $\mathcal{J}[\tilde{h},h]$, following standard methods
\cite{Janssen:1976,DeDominicis:1976,Janssen:1992}:\
\begin{equation}
\mathcal{J}[\tilde{h},h]=\gamma\int d^{d}x \, dt \Big\{  \tilde{h}\Big[
\gamma^{-1}\partial_{t}h+\partial_{\parallel}j_{\parallel}+\nabla_{\perp
}\mathbf{j}_{\perp}\Big]  -\tilde{h}^{2}\Big\}  \,. \label{def_J}%
\end{equation}
This has the advantage that both correlation and response functions can be
computed as appropriate functional averages, with statistical weight
$\exp(-\mathcal{J})$. The analysis can be simplified considerably if we
exploit the symmetries of $\mathcal{J}[\tilde{h},h]$, using the explicit forms
of the currents, Eq.~(\ref{curr}). First, the
symmetry $h(\mathbf{r},t)\rightarrow h(\mathbf{r},t)+a$ implies invariance
under a coordinate shift in the $z$-direction. Second, the theory is invariant
under tilts of the surface by an infinitesimal ``angle''\ $\mathbf{b}$, i.e.,
$h(\mathbf{r},t)\rightarrow h(\mathbf{r},t)+\mathbf{b}\cdot\mathbf{r}$
provided the tilt is accompanied by a transformation of the couplings, namely,
$\tau_{\parallel}\rightarrow\tau_{\parallel}-b_{\Vert}\lambda_{\parallel}$ and
$\tau_{\perp}\rightarrow\tau_{\perp}-b_{\Vert}\lambda_{\perp}$. Third,
particle conservation on the surface leads to invariance under the symmetry
transformation $\tilde{h}(\mathbf{r},t)\rightarrow\tilde{h}(\mathbf{r},t)+c$,
$h(\mathbf{r},t)\rightarrow h(\mathbf{r},t)+2c\gamma t$. Finally, we have a
symmetry under inversion, namely, $h(x_{\Vert},\mathbf{r}_{\bot}%
,t)\rightarrow-h(-x_{\Vert},\mathbf{r}_{\bot},t)\,$, $\tilde{h}(x_{\Vert
},\mathbf{r}_{\bot},t)\rightarrow-\tilde{h}(-x_{\Vert},\mathbf{r}_{\bot}%
,t)\,$. The most important of these
symmetries is the tilt invariance. Thanks to the associated Ward-Takahashi
identity \cite{Amit:1984,Zinn-Justin:1996}, the renormalizations of
$\tau_{\parallel}$ and $\tau_{\perp}$ can be related to those of
$\lambda_{\parallel}$ and $\lambda_{\perp}$, so that some exponent relations
will be valid to all orders in perturbation theory \cite{SPJ-Helsinki}.

Due to the anisotropy, there are two independent length scales. If only
\textit{parallel} lengths are rescaled, via $x_{\parallel}\rightarrow\alpha
x_{\parallel}$, the functional remains invariant provided $h\rightarrow
\alpha^{-1/2}h$, $\tilde{h}\rightarrow\alpha^{-1/2}\tilde{h}$ and
$\mu_{\parallel}\rightarrow\alpha^{4}\mu_{\parallel}$, $\tau_{\parallel
}\rightarrow\alpha^{2}\tau_{\parallel}$, $\mu_{\times}\rightarrow\alpha^{2}%
\mu_{\times}$ while $\lambda_{\parallel}\rightarrow\alpha^{7/2}\lambda
_{\parallel}$ and $\lambda_{\perp}\rightarrow\alpha^{3/2}\lambda_{\perp}$.
Likewise, if only \textit{transverse}\textrm{\ }lengths are rescaled, via
$\mathbf{r}_{\perp}\rightarrow\beta\mathbf{r}_{\perp}$, the functional remains
invariant provided $h\rightarrow\beta^{-(d-1)/2}h$, $\tilde{h}\rightarrow
\beta^{-(d-1)/2}\tilde{h}$ and $\mu_{\perp}\rightarrow\beta^{4}\mu_{\perp}$,
$\tau_{\perp}\rightarrow\beta^{2}\tau_{\perp}$, $\mu_{\times}\rightarrow
\beta^{2}\mu_{\times}$ while $\lambda_{\parallel}\rightarrow\beta
^{(d-1)/2}\lambda_{\parallel}$ and $\lambda_{\perp}\rightarrow\beta
^{(d+3)/2}\lambda_{\perp}$. As a result, the theory naturally gives rise to
effective expansion parameters which are invariant under these rescalings. The
precise forms of these parameters differ slightly for the four cases and will
be discussed next.

In addition to these purely spatial rescalings, we can perform a more general
dimensional analysis of Eq.~(\ref{def_J}), involving both spatial and temporal
degrees of freedom. It is standard to express it in terms of an external
length scale $\kappa^{-1}$. The key to the scaling of different terms in the
functional lies in the behavior of the control parameters $\tau_{\parallel}$
and $\tau_{\perp}$. Depending on whether they vanish or remain finite, the
Gaussian part of the dynamic functional is dominated by different terms in the
(infrared) limit of small momenta and frequencies.

{\it Case o:}  
If both $\tau_{\parallel}$ and $\tau_{\perp}$ are finite and
positive, the theory turns out to be purely Gaussian. Quartic derivatives can
be neglected in the infrared limit. It is natural to scale both parallel and
transverse momenta by $\kappa$, via $\left|  \mathbf{q}_{\bot}\right|
\propto$ $q_{\parallel}\propto\kappa$. As a result, time scales as
$\kappa^{-2}$ and the fields have dimensions $h(\mathbf{r},t)\propto
\kappa^{(d-2)/2}$ and $\tilde{h}(\mathbf{r},t)\propto\kappa^{(d+2)/2}$ so that
the nonlinear couplings scale as $\lambda_{\parallel}\propto\lambda_{\perp
}\propto\kappa^{-d/2}$ and are therefore irrelevant in any dimension $d>0$.
The resulting theory is a simple anisotropic generalization of the
Edwards-Wilkinson equation \cite{EdwardsWilkinson:1982},
\[
\gamma^{-1}\partial_{t}h=\tau_{\parallel}\partial_{\parallel}^{2}h+\tau
_{\perp}\nabla_{\perp}^{2}h+\eta\ .
\]
The anisotropies in the quadratic terms affect only nonuniversal amplitudes
and can be removed by a simple rescaling, without losing any information of
interest. As is well known, the two-point correlation function scales as
\[
C(\mathbf{r},t)=\left|  \mathbf{r}\right|  ^{2-d}c(t/\left|  \mathbf{r}%
\right|  ^{2})
\]
from which one immediately reads off the (isotropic) roughness exponent
$\chi=(2-d)/2$ and the dynamic exponent $z=2$. 
Since this case is so familiar (see \cite{Krug:1997} for a
detailed discussion), it does not need to be
considered further.

{\it Case i:}  
If $\tau_{\parallel}$ remains finite and positive but
$\tau_{\perp}$ vanishes, the two leading (Gaussian)\ terms in the dynamic
functional are $\mu_{\perp}\tilde{h}(\nabla_{\perp}^{2})^{2}h$ and
$\tau_{\parallel}\tilde{h}\partial_{\parallel}^{2}h$. This suggests that
parallel and transverse momenta scale differently, already at the Gaussian
level, namely, $\left|  \mathbf{q}_{\bot}\right|  \propto\kappa$ and
$q_{\parallel}\propto\kappa^{2}$. If we rewrite the scaling of parallel
momenta as $q_{\parallel}\propto\kappa^{1+\Delta}$, the anisotropic scaling
exponent $\Delta$ equals unity for the Gaussian theory. Time scales as
$\kappa^{-4}$, and $\tau_{\perp}\propto\kappa^{2}$ is strongly relevant.
$\mu_{\perp}$ can be set to $1$ by a transverse rescaling with an appropriate
$\beta,$ and $\mu_{\parallel}\propto$ $\mu_{\times}$ $\propto\kappa^{-2}$ are
strongly irrelevant (in the renormalization group sense). Introducing the
effective dimension $D=d+1$, one finds $h(\mathbf{r},t)\propto
\kappa^{(D-4)/2}$ and $\tilde{h}(\mathbf{r},t)\propto\kappa^{(D+4)/2}$. For
the nonlinear couplings, we obtain $\lambda_{\perp}\propto\kappa^{(4-D)/2}$
and $\lambda_{\parallel}\propto\kappa^{-D/2}$. Since $D$ is clearly positive,
the coupling $\lambda_{\parallel}$ becomes irrelevant. The upper critical
dimension $d_{c}$ for the theory is determined by $\lambda_{\perp}$, via
$0=4-D$ which leads to $d_{c}=3$. The invariant dimensionless effective
expansion parameter is $\tau_{\parallel}^{-3/4}\lambda_{\perp}\kappa
^{(d-3)/2}$ as shown by the rescaling $\lambda_{\perp}\rightarrow\alpha
^{3/2}\lambda_{\perp}$, $\tau_{\parallel}\rightarrow\alpha^{2}\tau_{\parallel
}$. At the Gaussian level, this case corresponds to a critical line
parametrized by $\tau_{\parallel}$. The mean-field values for the roughness
exponents are simple: In real space, one has $\chi_{\Vert}=\left(  3-d\right)
/4$ and $\chi_{\bot}=\left(  3-d\right)  /2$ while the momentum space
exponents are given by $\tilde{\chi}_{\Vert}=(2-d)/2$ and $\tilde{\chi}_{\bot
}=\left(  4-d\right)  /2$. While the potentially negative value of 
$\tilde{\chi}_{\Vert}$
might appear startling, it is simply a consequence of forcing Eq.~(\ref{SF_MS}%
) into the form of Eq.~(\ref{aniso_MS}).

{\it Case ii:} 
Here, $\tau_{\perp}$ remains finite while $\tau_{\parallel}$
vanishes. The Gaussian part of the functional is dominated by $\tilde
{h}\partial_{\parallel}^{4}h$ and $\tau_{\perp}\tilde{h}\nabla_{\perp}^{2}h$.
Again, even at the tree level, parallel and transverse momenta scale
differently: now, $q_{\parallel}\propto\kappa$ and $\left|  \mathbf{q}_{\bot
}\right|  \propto\kappa^{1+\Delta}$ with $\Delta=1$. Time scales as
$\kappa^{-4}$. The strongly relevant perturbation is $\tau_{\parallel}%
\propto\kappa^{2}$. One may still write $h(\mathbf{r},t)\propto\kappa
^{(D-4)/2}$ and $\tilde{h}(\mathbf{r},t)\propto\kappa^{(D+4)/2}$ but the
appropriate effective dimension is now $D\equiv(d-1)(1+\Delta)+1$. The two
nonlinearities switch roles so that $\lambda_{\parallel}\propto\kappa
^{(6-D)/2}$ and $\lambda_{\perp}\propto\kappa^{-D/2}$. In this case,
$\lambda_{\perp}$, $\mu_{\perp}$, and $\mu_{\times}$ are irrelevant while
$\lambda_{\parallel}$ becomes marginal at the upper critical dimension
$d_{c}=7/2$. The invariant dimensionless effective expansion parameter follows
from the rescalings as
$\mu_{\parallel}^{-7/8}\tau_{\bot}^{-(d-1)/4}\lambda_{\Vert}\kappa
^{(2d-7)/2}$. Again, it appears as if this case corresponds to a 
critical line parameterized by $\tau_{\perp}$. 
However, we will see below that the
order of the transition may well become first order once fluctuations are
included. We therefore refrain from quoting 
mean-field roughness exponents here.

{\it Case iii:}
Finally, we consider the multicritical point where both
$\tau_{\perp}$ and $\tau_{\parallel}$ vanish. Both momenta scale identically,
as $q_{\parallel}\propto\left|  \mathbf{q}_{\bot}\right|  \propto\kappa$, so
that $\Delta=0$ at the tree level. We choose $\beta$ so that $\mu_{\perp}$
scales to $1$. One obtains $\gamma t\propto$ $\kappa^{-4}$, $\tau_{\parallel
}\propto\tau_{\perp}\propto\kappa^{2}$, and $h(\mathbf{r},t)\propto
\kappa^{(D-4)/2}$, $\tilde{h}(\mathbf{r},t)\propto\kappa^{(D+4)/2}$, with
$D=d$. Both nonlinear couplings, $\lambda_{\parallel}$ and $\lambda_{\perp}$,
have the same upper critical dimension $d_{c}=6$. The effective expansion
parameters are $w\equiv\mu_{\times}/\sqrt{\mu_{\parallel}}$, $u_{\parallel
}\equiv\mu_{\parallel}^{-7/8}\lambda_{\parallel}\kappa^{(d-6)/2}$,
and\ $u_{\perp}\equiv\mu_{\parallel}^{-3/8}\lambda_{\perp}\kappa^{(d-6)/2}$.
In this case, the anisotropy exponent $\Delta$ vanishes at the tree level so
that, to this approximation, all four roughness exponents are equal, given by
$\left(  4-d\right)  /2$. We will see, however, that this changes already in
first order of perturbation theory.

In the following, we compute the scaling properties of correlation and
response functions for the physically most interesting case (i) in a one-loop
approximation. Our findings for cases (ii) and (iii) are reviewed only
briefly, leaving the full technical analysis to \cite{SPJ-Helsinki}.

\subsection{The one-loop approximation.}

We use dimensional regularization combined with minimal subtraction
\cite{Amit:1984,Zinn-Justin:1996}. The basic building blocks of the
perturbative analysis are the one-particle irreducible vertex functions
$\Gamma_{\tilde{N},N}(\{\mathbf{q},\omega\})$ with $\tilde{N}$ ($N$) $h$
($\tilde{h}$-) amputated legs. The notation $\{\mathbf{q},\omega\}$ is
short-hand for the full momentum- and frequency-dependence of these functions.
Focusing on the ultraviolet singularities, only those $\Gamma_{\tilde{N},N}$
with positive engineering dimension need to be considered. Taking into account
the symmetries and the momentum-dependence carried by the derivatives on the
external legs, the set of naively divergent vertex functions is reduced to
$\Gamma_{1,1}$ and $\Gamma_{1,2}$ which are computed to one-loop order. 
Some technical details are relegated to the Appendix.

\subsubsection{Case i: $\tau_{\perp}\rightarrow0$ and $\tau_{\parallel}>0$}

This is the simplest non-trivial case. Only one parameter, $\tau_{\perp}$,
needs to be tuned to access criticality. Since $\lambda_{\Vert}$ is
irrelevant, it may be set to zero. Neglecting all other irrelevant terms as
well, the functional simplifies to
\begin{eqnarray}
\mathcal{J}[\tilde{h},h]&=& \gamma\int d^{d}x dt\ \Big\{  \tilde{h}\bigl[
\gamma^{-1}\partial_{t}+(\nabla_{\perp}^{2})^{2}-\tau_{\parallel}%
\partial_{\parallel}^{2}-\tau_{\perp}\nabla_{\perp}^{2}\bigr]  h \nonumber \\
 &-&\tilde{h}%
^{2}-\lambda_{\perp}\tilde{h}(\partial_{\parallel}h)\nabla_{\perp}^{2}h\Big\}
\,. \label{case 1}%
\end{eqnarray}

Thanks to the momentum dependence of the nonlinear term, $\lambda_{\perp
}\tilde{h}(\partial_{\parallel}h)\nabla_{\perp}^{2}h$, all divergences in
$\Gamma_{1,1}$ and $\Gamma_{1,2}$ are already logarithmic and appear as simple
poles in $\varepsilon\equiv d_{c}-d$. In a minimal subtraction scheme, we
focus exclusively on these poles and their amplitudes to extract the
renormalizations. Since the nonlinearity is cubic in the field, the expansion
is organized in powers of $\lambda_{\perp}^{2}$; i.e., the first correction to
the tree level is always \textit{quadratic} for $\Gamma_{1,1}$ and
\textit{cubic} for $\Gamma_{1,2}$. The tilt invariance leads to a Ward
identity connecting $\Gamma_{1,1}$ and $\Gamma_{1,2}$, so that only the
divergences in $\Gamma_{1,1}$ need to be computed explicitly. Specifically,
the tilt transformation $h(\mathbf{r},t)\rightarrow h(\mathbf{r}%
,t)+\mathbf{b}\cdot\mathbf{r}$, $\tau_{\perp}\rightarrow\tau_{\perp}-b_{\Vert
}\lambda_{\perp}$ shows that the parameter $\mathbf{b}$ renormalizes as the
field $h$ itself. Hence, the term $\lambda_{\perp}h$ is renormalized by the
same factor as $\tau_{\perp}$.

Considering the perturbative contributions to $\Gamma_{1,1}(\mathbf{q}%
,\omega)$ further, we note that all of them carry external momenta, indicating
that the terms $\tilde{h}\partial_{t}h$ and $\gamma\tilde{h}^{2}$ do not
acquire any corrections. Moreover, particle conservation in conjunction with
invariance under parallel inversion and transverse rotations prevents the
emergence of corrections to the $\tau_{\perp}\tilde{h}\nabla_{\perp}^{2}h$
term, at any order in perturbation theory. Hence, we should expect only two
nontrivial renormalizations in $\Gamma_{1,1}$, namely those for the field $h$
and for the parameter $\tau_{\parallel}$. Leaving the detailed calculations of
the renormalized quantities to the Appendix, we seek an infrared stable fixed
point for the effective dimensionless coupling $u$, defined as
\begin{equation}
u\equiv A_{\varepsilon}\tau_{\parallel}^{-3/4}\kappa^{-\varepsilon/2}%
\lambda_{\perp}%
\end{equation}
where $A_{\varepsilon}$ is a simple geometric factor, defined in the Appendix.
A careful analysis of the flow equations for the renormalized parameters
reveals the presence of a single infrared stable fixed point, at
\begin{equation}
u^{\ast}=\pm4\sqrt{\frac{\varepsilon}{3}}\,\bigl( 1+O(\varepsilon)\bigr)
\ ,\label{eq:fps_case_tperp0}%
\end{equation}
where the sign is given by the sign of the initial coupling constant
$\lambda_{\perp}$. To obtain the scaling properties of correlation and
response functions, we exploit the fact that the bare theory is independent of
the external momentum scale $\kappa$. The resulting partial differential
equation (the renormalization group equation) and its solution 
is discussed in the Appendix. 
It predicts, specifically, the scaling form of the 
height-height correlation function, $C(\mathbf{r},t)$. Including
the critical parameter, $\tau_{\bot}$, in its list of arguments,
we find:
\label{G-scal_i}%
\begin{eqnarray}
C(\mathbf{r},t;\tau_{\bot})= l^{d+\Delta-4+\eta}
C(l^{1+\Delta}x_{\parallel},l\mathbf{r}_{\perp},l^{z}t;l^{-1/\nu_{\bot}}\tau_{\bot
})\,\ \label{G_i}
\end{eqnarray}
where $l$ is an arbitrary flow parameter. This behavior of $C(\mathbf{r},t)$
is clearly a natural anisotropic generalization
\cite{SchmittmannZia:1995} of the usual scaling form of critical dynamics. 
Moreover, it is clearly consistent with the anticipated scaling behavior,
Eq.~(\ref{G_RS}), especially if we set $\tau_{\bot}=0$. 
The exponents $\nu_{\bot}$, $\eta$, and $z$ have their usual 
meanings: $\nu_{\bot}$ controls
the scaling of the strongly relevant coupling $\tau_{\bot}$, $\eta$ is the
anomalous dimension of the field and controls critical correlation functions,
and $z$ is the dynamic exponent, relating spatial and temporal fluctuations at
criticality. $\Delta$ is the strong anisotropy exponent introduced in Section
III. Exploiting the symmetries of the theory fully (see Appendix), we find
that $\nu_{\bot}$, $\eta$, and $z$ are 
related by scaling laws which are exact (at
least within perturbation theory):
\[
z=4-\eta\,\qquad\text{and}\qquad1/\nu_{\bot}=2-\eta\text{ }.
\]
Assuming that $u^{\ast}$ remains nonzero at higher orders of perturbation
theory, its flow equation gives us another exact scaling law, relating
$\Delta$ and $\eta$:\
\[
\Delta+\eta=2-d/3\,
\]
As a consequence, only a \textit{single} exponent, e.g., $\eta$, has to be
computed order by order in perturbation theory. Our one-loop calculation
results in
\begin{equation}
\eta=-2\varepsilon/3+O(\varepsilon^{2})\,. \label{eta-i}%
\end{equation}
Now, all others follow from exponent identities which are exact, at least
as long as there are no non-perturbative corrections. 

\subsubsection{Case ii: $\tau_{\parallel}\rightarrow0$ and $\tau_{\perp}>0$}

\label{case_tpar0} This is the second non-trivial case. Neglecting irrelevant
terms, the dynamic functional simplifies to
\begin{align}
\mathcal{J}[\tilde{h},h]= & \gamma\int d^{d}x\, dt\ \Big\{  \tilde{h}\Big[
\gamma^{-1}\partial_{t}+\mu_{\parallel}\partial_{\parallel}^{4}-\tau
_{\parallel}\partial_{\parallel}^{2}-\tau_{\perp}\nabla_{\perp}^{2}\Big]h
\nonumber \\ 
& -\tilde{h}^{2}-\lambda_{\parallel}\tilde{h}(\partial_{\parallel}%
h)\partial_{\parallel}^{2}h\Big\}  \,. \label{case 2}%
\end{align}
and the upper critical dimension is now $d_{c}=7/2$.

Following the same approach as in the previous case, we first seek an infrared
stable fixed point. The existence of such a fixed point guarantees that the
model exhibits a scale-invariant regime where roughness exponents can be
defined. As before, we define a suitable effective coupling $u\propto
\mu_{\parallel}^{-7/8}\tau_{\perp}^{-(d-1)/4}\lambda_{\parallel}%
\kappa^{(2d-7)/2}$ and analyze its flow. In this case, however, we
find no stable fixed points corresponding to physically meaningful (i.e.,
real) values of the coupling $u$, at least to this order in perturbation
theory \cite{SPJ-Helsinki}. The lack of such fixed points often 
indicates a first order transition, but only a more detailed
analysis of the underlying mean-field theory or a careful computational study
will resolve this issue.

\subsubsection{Case iii: $\protect\tau _{\parallel }\rightarrow 0$ and 
$\protect\tau_{\perp }\rightarrow 0$}

Finally, we briefly review our results for the multicritical point where both
critical parameters vanish simultaneously. This case was previously
studied by Marsili et al.\cite{MarsiliETAL:1996}, 
in a momentum shell decimation scheme. In this procedure, a hard momentum
cutoff prevents the emergence of ultraviolet divergences in the 
momentum integrals. However, in theories with strong
anisotropy and nonlinearities carrying multiple derivatives, 
the perturbative corrections depend on how the cutoff is implemented,
requiring extreme care.  
This may explain why we were unable to reproduce the earlier 
\cite{MarsiliETAL:1996} results. In contrast, our findings
are easy to check, since our field-theoretic
approach does not suffer from these complications. 

The full functional, defined by Eqs.~(\ref{def_J}) and (\ref{curr}), 
now comes into play. To ensure the stability of the
critical theory at the tree level, we demand $\mu_{\parallel}\,q_{\parallel
}^{4}+2\mu_{\times}q_{\parallel}^{2}\,\mathbf{q}_{\bot}^{2}+(\mathbf{q}_{\bot
}^{2})^{2}\geq0$. This limits the physical range of $\mu_{\parallel}$ and
$\mu_{\times}$ to $\mu_{\parallel}>0$ and $\mu_{\times}>-\sqrt{\mu_{\parallel
}}$. To complicate matters further, both nonlinear couplings, $\lambda
_{\parallel}$ and $\lambda_{\perp}$, are marginal at the upper critical
dimension $d_{c}=6$. While the detailed calculations become more involved, the
technical analysis remains straightforward \cite{SPJ-Helsinki}. In particular,
thanks to the Ward identity, all renormalizations can still be obtained from
the two-point function, $\Gamma_{1,1}$.

The key results are as follows. Again, we define appropriate effective
couplings. One of these,
\[
w\equiv\frac{\mu_{\times}}{\sqrt{\mu_{\parallel}}}%
\]
appears in the propagator and generates $w$-dependent algebraic coefficients
in the perturbation expansion. The other two,
\begin{eqnarray}
u_{\Vert} &\equiv& C_{\varepsilon}\mu_{\parallel}^{-7/8}(1+w)^{-5/4}%
\lambda_{\Vert}\kappa^{(d-6)/2}\,~, \nonumber \\
u_{\bot}&\equiv& C_{\varepsilon}%
\mu_{\parallel}^{-3/8}(1+w)^{-5/4}\lambda_{\bot}\kappa^{(d-6)/2}%
\nonumber
\end{eqnarray}
control the nonlinar terms and are treated order by order in perturbation
theory. $C_{\varepsilon}$ just absorbs some common geometric\ constants. A
careful analysis of the flow equations for these three couplings shows
\cite{SPJ-Helsinki} that there is only a single, physically meaningful
infrared stable fixed point, given by
\begin{eqnarray}
w^{\ast}&=&2\sqrt{\frac{3}{5}}-1+O(\varepsilon)\,,\qquad 
u_{\bot}^{\ast}=0 \,, \nonumber \\
u_{\Vert}^{\ast}&=&\pm\sqrt{\frac{7\sqrt{15}+25}{11}\,}\varepsilon^{1/2}
+O(\varepsilon
^{3/2})\,. \label{fp_4}%
\end{eqnarray}
This contradicts earlier results\cite{MarsiliETAL:1996} where a
fixed point with $u_{\bot}^{\ast}\neq0$ was supposedly found.

Exploiting the symmetries of our theory fully and assuming that $u_{\bot
}^{\ast}=0$ remains valid to all orders in perturbation theory
\cite{SPJ-Helsinki}, we can determine the critical exponents associated with
this model. Remarkably, we find that only a single exponent must be computed
\textit{explicitly} within the $\varepsilon$-expansion, e.g., $\Delta$. The
scaling behavior of the height-height correlation function obeys the general
form given in Eq.~(\ref{G_i}), except that both critical parameters now
appear, each with its own scaling exponent, $\nu_{\bot}$ and $\nu_{\Vert}$,
respectively. Again, this confirms the anticipated scaling,
Eq.~(\ref{G_RS}). The exponents, however, take different values here, 
demonstrating that cases (i) and (iii) fall into distinct universality
classes.  
Our one-loop calculation yields
\begin{equation}
\Delta=\frac{23+6\sqrt{15}}{11}\varepsilon+O(\varepsilon^{2})\,
\end{equation}
The remaining exponents $\eta$, $z$ and $\nu_{\perp}$ do not acquire any
corrections beyond the tree results, so that
\[
\eta=0\ ,\qquad z=4\ ,\qquad\nu_{\perp}=\frac{1}{2}%
\]
to all orders in perturbation theory. The exponent $\nu_{\parallel}$ is
related to $\Delta$ through an exact scaling relation, namely,
\begin{equation}
\nu_{\Vert}=\frac{2}{d-2+3\Delta}\,.\label{nu_par_iii}%
\end{equation}
These results are sufficient to evaluate the associated roughness exponents.

\section{Results for the roughness exponents}

To begin with, we recall that the scaling forms of the two-point correlations
for cases (i) and (iii) are indeed consistent with Eq.~(\ref{G_RS}).  
We may therefore immediately express the roughness exponents, 
Eqs.~(\ref{Chi_RS}) and (\ref{Chi_MS}), in terms of the exponent $\Delta$ and
$\eta$ for the two universality classes.

For case (i), characterized by $\tau_{\bot}\rightarrow0$ at positive 
$\tau_{\Vert}$,
we found only one independent exponent, namely  
$\eta=-2\varepsilon/3+O(\varepsilon^{2})$ with $\Delta=2-d/2-\eta$. Writing all
four roughness exponents in terms of $\eta$, we arrive at expressions which are 
exact to all orders in $\varepsilon=d-3$:
\begin{subequations}
\begin{align}
\chi_{\bot}  & =1-d/3\, & \chi_{\Vert}&  =\frac{1-d/3}{3-d/3-\eta}
\, \nonumber \\
\tilde{\chi}_{\bot}  &=(4-d-\eta)/2\,&  \tilde{\chi}_{\Vert}  
&=\frac{4-\eta}{3-d/3-\eta}-\frac{d}{2}\,\,. \nonumber%
\end{align}
\end{subequations}
The mean-field values are easily recovered by setting $\varepsilon=0$. The
physically most interesting case corresponds to a surface grown on a
two-dimensional substrate, i.e., $d=2$ and $\varepsilon=1$. For this
situation, one obtains $\chi_{\bot}=1/3$ and $\chi_{\Vert}=1/9+O(\varepsilon
^{2})$ while $\tilde{\chi}_{\Vert}=-2/9+O(\varepsilon^{2})$ and $\tilde{\chi%
}_{\bot}=4/3+O(\varepsilon^{2})$. Remarkably, the exponent $\chi_{\bot}=1/3$
is actually \textit{exact}, at least to all orders in perturbation theory.

At the multicritical point, i.e., case (iii) with $\tau_{\bot}\rightarrow0,
\tau_{\Vert}\rightarrow0$, we found $\eta=0$ and a nontrivial 
$\Delta$. Hence, 
\begin{subequations}
\begin{align}
\chi_{\bot}  &  =\frac{1}{2}[4-(d+\Delta)] & \chi_{\Vert}  &  
=\frac{4-(d+\Delta)}{2(1+\Delta)} \nonumber \\
\tilde{\chi}_{\bot}  &  =\frac{1}{2}(4-d) & \tilde{\chi}_{\Vert}  &  
=\frac{2}{1+\Delta}-\frac{d}{2} \,. \nonumber%
\end{align}
\end{subequations}
All of them are negative near the upper critical dimension $d_{c}=6$. In order
to access the physical ($d=2$) situation, one has to set $\varepsilon=4$ here
which gives a huge anisotropy exponent, $\Delta\simeq16.814$, if one naively
uses the one-loop result. While roughness exponents can in principle be
calculated, we should not place much confidence in their numerical values.
Nevertheless, we still obtain testable predictions for this case, 
namely, the general scaling form of the height-height correlations as well 
as the scaling laws relating different exponents. 

Needless to say, in the absence of an infrared stable fixed point, the
renormalization group gives us no information about possible roughness
exponents for case (ii)\ ($\tau_{\perp}>0$, $\tau_{\parallel}\rightarrow0$).
In fact, if the scenario of a first order transition line applies, the whole
concept of scaling exponents would be misplaced.

\section{Conclusions}

To summarize, we have analyzed the long-time, large distance scaling
properties of a surface growing under ideal MBE-type conditions, subject to an
incident particle beam tilted away from the surface normal. This selects a
particular (``parallel'')\ direction in the substrate plane, so that the
resulting growth equations for the (single-valued) height field are spatially
anisotropic. In particular, an \textit{effective} surface tension 
becomes anisotropic,
contributing the terms $\tau_{\parallel}\partial_{\parallel}^{2}h+\tau_{\bot
}\nabla_{\bot}^{2}h$ to the right hand side of the Langevin equation. If both
of these parameters, $\tau_{\parallel}$ and $\tau_{\perp}$, are positive, the
surface is described by the Edwards-Wilkinson model. However, depending on
experimental control parameters such as temperature, incident flux rate and
angle, or particle size, either $\tau_{\parallel}$ or $\tau_{\perp}$ or both
can vanish, generating significantly different surface properties. Each of
these three possibilities leads to a distinct field theory with different
upper critical dimension. We find $d_{c}=3$ if $\tau_{\perp}\rightarrow0$ at
finite $\tau_{\parallel}$, $d_{c}=7/2$ if $\tau_{\parallel}\rightarrow0$ at
finite $\tau_{\perp}$, and $d_{c}=6$ if both vanish. Only the first and the
third case lead to scale-invariant behavior; the second one may in fact
trigger a first-order phase transition whose properties lie outside the scope
of our RG techniques. Focusing on the first and third case, we find two
distinct, novel surface universality classes. For both, we compute the scaling
behavior of the two-point height-height correlation function and carefully
extract four different roughness exponents. Two of these characterize the
height fluctuations of the surface in real space, scanned either along the 
parallel or the transverse directions; 
the remaining two characterize scattering data
with momentum transfer either along the parallel or 
the transverse directions. When
analyzing experimental data, care must be taken in identifying 
the correct exponent.

Clearly, the third case requires the careful tuning of two parameters,
$\tau_{\parallel}$ and $\tau_{\perp}$. If we can substantiate the presence of
a first-order line for $\tau_{\parallel}\rightarrow0$ at finite $\tau_{\perp}%
$, the point $\tau_{\parallel}=\tau_{\perp}=0$ would in fact be a critical
endpoint, since it separates a line of second order transitions from a line of
first order ones. In order to access it in a typical experiment, at least two
control parameters have to be tuned very carefully, making it difficult to
observe. For this reason, either the second or the first order lines should be
more accessible experimentally. From the RG perspective, even if 
$\tau_{\parallel}$ is set to zero initially, it will be generated 
under RG transformations, resulting in a non-zero value; 
this is not the case for the $\tau_{\perp}\nabla_{\perp}^{2}h$ 
contribution. In that sense, we believe that the most
physically relevant theory (apart from Edwards-Wilkinson behavior) is the one
with $\tau_{\parallel}>0$ and $\tau_{\perp}=0$. Amongst our key results for
this model are the roughness exponents for real-space surface scans. For the
physically most interesting case of a two-dimensional surface, we find
$\chi_{\bot}=1/3$ if the fluctuations are measured along the transverse
direction, and $\chi_{\Vert}=1/9+O(\varepsilon^{2})$ for scans along the
parallel direction. While the value for $\chi_{\Vert}$ may be modified by
higher-order contributions in perturbation theory, the result for $\chi_{\bot
}$ is exact. 

Finally, we turn to possible experimental evidence for these exponents. 
Thin films of vapor-deposited gold on smooth glass
surfaces \cite{HerrastiETAL:1992,SalvarezzaETAL:1992} were previously 
proposed \cite{MarsiliETAL:1996} as possible
realizations of our theory. The deposit surfaces were imaged by scanning
tunneling microscopy (STM), and their mean-square width was measured as a
function of STM\ scan length, resulting in a roughness exponent of $0.35$
\cite{HerrastiETAL:1992,SalvarezzaETAL:1992}. The temperature is sufficiently
low ($T=298$ K) so that desorption is negligeable. Even at small angles of
incidence (between $2^{o}$ and $25^{o}$), the growth is
anisotropic. The STM images \cite{HerrastiETAL:1992} show some 
evidence for striped pattern formation,
if larger sample areas ($\sim3\times 10^3$ nm on each side) are imaged. 
It is encouraging that a growth exponent close to 1/3 is observed since this
matches our prediction for the transverse direction; 
unfortunately, there seems to be no evidence for the much smaller exponent
($1/9$) which should control the parallel direction. 
A final resolution of these issues has to await a more detailed analysis of the
experimental data.

\acknowledgments   We thank U.C. T\"{a}uber, R.K.P. Zia, J. Krug, A. Hartmann,
and E. Yewande for helpful discussions. This work is partially supported by
NSF through DMR-0308548 and DMR-0414122. GP acknowledges the Alexander von
Humboldt foundation for their support.

\appendix         

\section*{Appendix}

In the following, some technicalities associated with case (i) are presented.
We will switch freely between the $\left(  \mathbf{q},t\right)  $ and the
$\left(  \mathbf{q},\omega\right)  $ representations, depending on which one
is more convenient. We first collect the elements of perturbation theory and
then discuss the one-loop corrections. Neglecting all nonlinearities\ in 
Eq.~(\ref{case 1}) allows us to identify the bare propagator, $G_{0}%
(\mathbf{q},t)\,$, and the bare correlator, $C_{0}(\mathbf{q},t)\,$, via
\begin{eqnarray}
G_{0}(\mathbf{q},t)\,\delta(\mathbf{q}-\mathbf{q}^{\prime}) &\equiv& \langle
h(\mathbf{q},t)\tilde{h}(-\mathbf{q}^{\prime},0)\rangle_{0} \nonumber \\
&=& \theta(t)\,\exp\bigl( -\Gamma(\mathbf{q})\gamma t\bigr) \,,\nonumber \\
C_{0}(\mathbf{q},t)\,\delta(\mathbf{q}-\mathbf{q}^{\prime}) &\equiv& \langle
h(\mathbf{q},t)h(-\mathbf{q}^{\prime},0)\rangle_{0} \nonumber \\
&=& \Gamma(\mathbf{q})^{-1}G_{0}(\mathbf{q},|\,t|)\,,
\end{eqnarray}
where
\begin{equation}
\Gamma(\mathbf{q})\equiv\mu_{\parallel}q_{\parallel}^{4}+\tau_{\parallel
}q_{\parallel}^{2}+\tau_{\perp}\mathbf{q}_{\bot}^{2} \nonumber
\end{equation}
Here, $\langle\cdot\rangle_{0}$ denotes a functional average with Gaussian
($\lambda_{\perp}=0$)\ weight. The Heaviside function $\theta(t)$ is defined
with $\theta(0)=0$. Turning to the interaction terms, it is convenient to
rewrite them in a symmetrized form. In Fourier space, the expression for the
three-point vertex reads
\begin{equation}
V(\mathbf{q}_{1},\mathbf{q}_{2},\mathbf{q}_{3})\equiv i\gamma\lambda_{\perp
}\Big[ \mathbf{q}_{1\bot}\cdot\bigl( q_{2\parallel}\,\mathbf{q}_{3\bot
}+\mathbf{q}_{2\bot}\,q_{3\parallel}\bigr) -q_{1\parallel}\,\mathbf{q}_{2\bot
}\cdot\mathbf{q}_{3\bot}\Big] \,\label{vertex}%
\end{equation}
with $\mathbf{q}_{1}+\mathbf{q}_{2}+\mathbf{q}_{3}=0$. The sign convention is
such that all momenta attached to a vertex are incoming.

We are using dimensional regularization, so that ultraviolet divergences
appear as simple poles in $\varepsilon\equiv d_{c}-d$, with $d_{c}=3$. In a
minimal subtraction scheme, we focus exclusively on these poles and extract
the renormalization constants from their amplitudes. Thanks to the
symmetries of the theory, we
anticipate only two nontrivial renormalizations, namely those for the field
$h$ and for the parameter $\tau_{\parallel}$, both of which can be obtained
from $\Gamma_{1,1}$. To one loop order, the explicit expression for the
singular part of the two-point vertex function is given by:
\begin{eqnarray}
\Gamma_{1,1}(\mathbf{q},\omega)_{\mathrm{pole}} &=& i\omega+\gamma\big[
\tau_{\parallel}q_{\parallel}^{2}+(\mathbf{q}_{\bot}^{2})^{2}+\tau_{\perp
}\mathbf{q}_{\bot}^{2}\big]  \\
&+& \gamma\frac{u^{2}}{8\varepsilon}\big[
2\tau_{\parallel}q_{\parallel}^{2}-(\mathbf{q}_{\bot}^{2})^{2}\big]
+O(u^{4})\,,\nonumber
\end{eqnarray}
where the effective expansion parameter $u$ is given by
\begin{equation}
u\equiv A_{\varepsilon}\tau_{\parallel}^{-3/4}\kappa^{-\varepsilon/2}%
\lambda_{\perp}%
\end{equation}
and $A_{\varepsilon}$ summarizes a geometric factor which appears in all
Feynman diagrams:
\[
A_{\varepsilon}\equiv\frac{S_{d-1}}{(2\pi)^{d}}\sqrt{\pi}\Gamma\left(
\frac{1-\varepsilon}{2}\right)  \Gamma\left(  \frac{1+\varepsilon}{2}\right)
\]
$S_{d}$ is the surface area of the $d$-dimensional unit sphere.

Keeping in mind that there are only two nontrivial renormalizations to all
orders, we introduce renormalized quantities via
\begin{eqnarray*}
h  \rightarrow\mathring{h}&=&Z^{1/2}h \quad \,\,\,\, 
\tilde{h}\rightarrow\mathring
{\tilde{h}}=Z^{-1/2}\tilde{h}\quad 
\gamma\rightarrow\mathring{\gamma}=Z\gamma 
\nonumber \\
\tau_{\perp}  \rightarrow\mathring{\tau}_{\perp}&=& Z^{-1}\tau_{\perp}%
\quad \,  \tau_{\parallel}\rightarrow\mathring{\tau}_{\parallel}=Z^{-1}Z_{\tau
}\tau_{\parallel} \nonumber \\
\lambda_{\perp}  \rightarrow\mathring{\lambda}_{\perp}&=&Z^{-3/2}%
\lambda_{\perp} \nonumber
\end{eqnarray*}
The renormalized vertex function $\Gamma_{1,1}$ is defined by demanding that
\begin{equation}
\Gamma_{1,1}(\mathbf{q},\omega,\gamma,\tau_{\perp},\tau_{\parallel}%
,u,\kappa)\equiv\mathring{\Gamma}_{1,1}(\mathbf{q},\omega,\mathring{\gamma
},\mathring{\tau}_{\perp},\mathring{\tau}_{\parallel},\mathring{\lambda
}_{\perp})
\end{equation}
be pole-free. One finds%
\begin{align}
Z &  =1+\frac{u^{2}}{8\varepsilon}+O(u^{4})\,,\nonumber\\
Z_{\tau} &  =1-\frac{u^{2}}{4\epsilon}+O(u^{4})\,. \label{res_Z_2}
\end{align}
The corresponding Wilson functions are defined as the logarithmic derivatives
of the associated $Z$-factors, at constant bare quantities, i.e.,
\begin{eqnarray}
\zeta &\equiv& \kappa\partial_{\kappa}\ln Z|_{\text{bare}}=-\frac{u^{2}}%
{8}+O(u^{4})\ ,\qquad \nonumber \\
\zeta_{\tau} &\equiv& \kappa\partial_{\kappa}\ln Z_{\tau
}|_{\text{bare}}=\frac{u^{2}}{4}+O(u^{4})\label{def_gammas}%
\end{eqnarray}
The flow of the dimensionless effective coupling constant $u$ under
renormalization is controlled by the Gell-Mann--Low function,
\begin{align}
\beta(u) &  \equiv\kappa\partial_{\kappa}u|_{\text{bare}}=u\big[
-\frac{\varepsilon}{2}+\frac{3}{4}\left(  \zeta+\zeta_{\tau}\right)  \big]
\,\nonumber\\
&  =u\big[ -\frac{\varepsilon}{2}+\frac{3}{32}u^{2}+O(u^{4})\big]
\,.\label{beta_i}%
\end{align}

The renormalization group equation (RGE) for the Green functions,
i.e., the connected correlation functions with $N$ ($\tilde{N}$)
external $h$ ($\tilde{h}$) legs, simply states that the bare theory is
independent of the external momentum scale $\kappa$:
\begin{eqnarray*}
0&=&\kappa\frac{d}{d\kappa}\mathring{G}_{N,\tilde{N}}(\{\mathbf{r}%
,t\};\mathring{\tau}_{\bot},\mathring{\tau}_{\Vert},\mathring{\gamma
},\mathring{\lambda}_{\perp}) \label{RGE-i} \\
&=& \kappa\frac{d}{d\kappa}Z^{(N-\tilde{N})/2}%
G_{N,\tilde{N}}(\{\mathbf{r},t\};\tau_{\bot},\tau_{\Vert};u,\gamma
,\kappa)\,.\nonumber%
\end{eqnarray*}
In our case, this equation takes the form
\begin{eqnarray}
\Big[ \kappa\frac{\partial}{\partial\kappa}&+&\beta\frac{\partial}{\partial
u}-\zeta\gamma\frac{\partial}{\partial\gamma}+\zeta\tau_{\perp}\frac{\partial
}{\partial\tau_{\perp}}+ \nonumber \\
&+&(\zeta-\zeta_{\tau})\tau_{\parallel}\frac{\partial
}{\partial\tau_{\parallel}} 
+\frac{\zeta}{2}(N-\tilde{N})\Big] G_{N,\tilde{N}}%
=0\,.\label{RGE-ii}%
\end{eqnarray}
Asymptotic scaling results from this partial differential equation at an
infrared stable fixed point that is a solution of $\beta(u^{\ast})=0$ with
$\beta^{\prime}(u^{\ast})>0$. The one-loop approximation Eq.~(\ref{beta_i})
leads to the stable fixed point
\begin{equation}
u^{\ast}=\pm4\sqrt{\frac{\varepsilon}{3}}\,\bigl( 1+O(\varepsilon)\bigr)
\ ,\label{fp}%
\end{equation}
where the sign is given by the sign of the initial coupling constant
$\lambda_{\perp}$. Thus, $u^{\ast}$ is non-zero in the $\varepsilon
$-expansion, and making the reasonable assumption that this remains true to
all orders in perturbaton theory, we find from Eq.~(\ref{beta_i}) the exact
relation
\begin{equation}
\zeta^{\ast}+\zeta_{\tau}^{\ast}=\frac{2\varepsilon}{3}\label{sc-i}%
\end{equation}
where we have denoted the Wilson $\zeta$-functions, evaluated at the fixed
point, by a superscript $\ast$. Eq.~(\ref{RGE-ii})  can be solved easily at
the fixed point $u^{\ast}$, using the method of characteristics, combined with
dimensional analysis and the rescaling invariances. If we suppress 
unnecessary arguments, the solution can be written in the form
\begin{equation}
G_{N,\tilde{N}}(\{\mathbf{r},t\};\tau_{\bot})=l^{\delta_{N,\tilde{N}}%
}G_{N,\tilde{N}}(\{l^{1+\Delta}x_{\parallel
},l\mathbf{r}_{\perp},l^{z}t\};l^{-1/\nu}\tau_{\bot})\, \nonumber
\end{equation}
where $l$ is an arbitrary flow parameter. The two critical exponents
$\Delta$ and $\eta$
can be expressed in terms of the Wilson functions as 
\begin{equation}
\Delta=1+\frac{\zeta_{\tau}^{\ast}-\zeta^{\ast}}{2}~,\qquad\eta\,=\zeta^{\ast
} \label{Delta}
\end{equation}
while scaling laws give us the remaining exponents $\nu$, $z$, and the 
overall scaling exponent $\delta_{N,\tilde{N}}$: 
\begin{eqnarray}
\label{exp_i}%
z&=&4-\eta\,,\qquad\qquad1/\nu=2-\eta \,,    \label{scal_laws} \\
\delta_{N,\tilde{N}}&=&\frac{N}{2}(d+\Delta-4+\eta)
+\frac{\tilde{N}}{2}(d+\Delta+4-\eta) \nonumber%
\end{eqnarray}
Finally, Eq.~(\ref{sc-i}) provides another exact scaling law 
relating $\Delta$ and $\eta$, provided $u^{\ast}\neq0$ holds to all orders:
\[
\Delta+\eta=2-d/3\,
\]
As a consequence, only a \textit{single} exponent, e.g., $\eta$, has to be
computed order by order in perturbation theory, and all others follow from
exponent identities.

\end{document}